# Airy-Gaussian vortex beams in the fractional nonlinear-Schrödinger medium

Shangling He[1,7], Kangzhu Zhou[2,7], Boris A. Malomed[3,4], Dumitru Mihalache[5], Liping Zhang[1], Jialong Tu[1], You Wu[1], Jiajia Zhao[1], Xi Peng[6], Yingji He[6], Xiang Zhou[1], and Dongmei Deng[1]*

*[1]Guangdong Provincial Key Laboratory of Nanophotonic Functional Materials and Devices, South China Normal University, Guangzhou 510631, China*

*[2]School of Electronic and Computer Engineering, Peking University, Shenzhen, Guangdong 518000, China*

*[3]Department of Physical Electronics, School of Electrical Engineering, Faculty of Engineering, and Center for Light-Matter Interaction, Tel Aviv University, Tel Aviv 69978, Israel*

*[4]Instituto de Alta Investigación, Universidad de Tarapacá, Casilla 7D, Arica, Chile*

*[5]Horia Hulubei National Institute for Physics and Nuclear Engineering, P.O. Box MG-6, RO-077125, Bucharest-Magurele, Romania*

*[6]School of Photoelectric Engineering, Guangdong Polytechnic Normal University, Guangzhou 510665, China*

*\*Corresponding author: dmdeng@263.net*

*[7]These authors contributed equally*

## Abstract

We address the propagation of vortex beams with the circular Airy-Gaussian shape in a (2+1)-dimensional optical waveguide modeled by the fractional nonlinear Schrödinger (FNS) equation. Systematic analysis of autofocusing of the beams reveals a strongly non-monotonous dependence of the peak intensity in the focal plane on the corresponding Lévy index $\alpha$, with a strong maximum at $\alpha \approx 1.4$. Effects of the nonlinearity strength, the ratio of widths of the Airy and Gaussian factors in the input, as well as the beams' vorticity, on the autofocusing dynamics are explored. In particular, multiple autofocusing events occur if the nonlinearity is strong enough. Under the action of the azimuthal modulational instability, an axisymmetric beam may split into a set of separating bright spots. In the case of strong fractality (for $\alpha$ close to 1), the nonlinear beam self-traps, after the first instance of autofocusing, into a breathing vortical quasi-soliton. Radiation forces induced by the beams' field are considered too, and a capture position for a probe nanoparticle is thus identified.

## 1. Introduction

The concept of fractals in physics was introduced by B. B. Mandelbrot in 1982 [1]. Then, much interest has been drawn to fractional quantum mechanics, which is produced by the Feynman path integrals applied to orbits of classical particles in the form of Lévy flights, instead of the ordinary Brownian motion [2,3]. The wave function representing such orbits obeys the fractional Schrödinger equation (FSE) [4], which can be realized in various physical contexts [5-9]. In the context of the classical wave propagation, the FSE producing dual Airy beams in a spherical optical cavity was introduced by Longhi [10]. The FSE was then extended by including external potentials and nonlinear terms [11-13]. Concerning such settings, it is relevant to mention that, in the presence of a longitudinal modulation potential, the period of Rabi oscillations and the efficiency of the resonant conversion in the fractional medium can be controlled by the respective Lévy index (LI) [14]. The band structure produced by the FSE which includes a *PT*-symmetric potential was studied too [15].

Recently, much interest has been drawn to optical solitons of the fundamental and vortex types, produced by the fractional nonlinear Schrödinger (FNS) equation [16-25]. The interest is driven by the natural possibility to observe nonlinear effects in optical cavities and waveguides. As usual [26], stability of vortex solitons carrying the topological charge is a crucial issue. In particular, it was found that such states, supported by an underlying photonic lattice, are stable in a certain interval of values of the propagation constant, depending on the underlying LI [24].

A class of physically relevant axisymmetric modes is represented by ring-shaped Airy beams in nonlinear media [27-31], including circular Airy-Gaussian beams with embedded vorticity [32], autofocusing beams with a spiral phase [33], and Airy beams featuring periodic self-imaging under the action of a potential barrier [34]. However, to the best of our knowledge, the propagation and autofocusing of Airy-Gaussian vortex beams (AGVBs) were not yet studied in the framework of FNS media. Here, we address this setting; in particular, the results are compared to those produced by the propagation governed by the usual nonlinear Schrödinger equation.

The paper is structured as follows. The model is formulated in Section 2. The autofocusing dynamics featured by AGVBs is considered in Section 3, where the influence of parameters on the dynamics are considered in detail. Several effects are observed under the action of the nonlinearity, *viz.*, multiple self-focusing, self-trapping of quasi-solitons in the case of strong fractality, and splitting of axisymmetric beams under the action of the azimuthal modulational instability. The paper is concluded by Section 4.

## 2. The model

We consider the propagation of AGVBs governed by the cubic FNS equation [35, 36]

$$i\frac{\partial u}{\partial z} - \frac{1}{2kw^{2-\alpha}}\left(-\frac{\partial^2}{\partial x^2} - \frac{\partial^2}{\partial y^2}\right)^{\alpha/2}u + \frac{n_2 k}{n_0}|u|^2 u = 0, \qquad (1)$$

where $u$ is the local amplitude of the optical wave, $z$ is the propagation distance, $w$ is a characteristic transverse scale, which determines the respective Rayleigh length $Z_R = kw^2/2$, $k = 2\pi/\lambda$ corresponds to carrier wavelength $\lambda$, $\alpha$ is the LI ($1 < \alpha \leq 2$), $x$ and $y$ are scaled transverse coordinates, $n_0$ is the background refractive index, and $n_2 > 0$ is the Kerr coefficient. The fractional-diffraction operator in Eq. (1) is defined by the known integral expression [2,4,5],

$$\left(-\frac{\partial^2}{\partial x^2} - \frac{\partial^2}{\partial y^2}\right)^{\alpha/2}u(x,y,z) = \frac{1}{(2\pi)^2}\iint dq_x\,dq_y\,\exp\left(iq_x x + iq_y y\right)\left(q_x^2 + q_y^2\right)^{\alpha/2}\hat{u}\left(q_x, q_y, z\right), \quad (2)$$

where the two-dimensional Fourier transform is

$$\hat{u}\left(q_x, q_y, z\right) = \iint dx'dy'\exp\left(-iq_x x' - iq_y y'\right)u\left(x', y', z\right). \qquad (3)$$

If stationary localized solutions to Eq. (1) (solitons) with a real propagation constant $K > 0$ are looked for as

$$u\left(x, y, z\right) = e^{iKz}U_K\left(x, y\right), \qquad (4)$$

a straightforward corollary of Eq. (1) is a scaling relation for the total power [37]:

$$P(K) = \iint\left|U_K\left(x, y\right)\right|^2 dxdy = \text{const}\cdot K^{1-2/\alpha}. \qquad (5)$$

It is relevant to mention that, in the case of $\alpha < 2$, relation (5) does not satisfy the well-known necessary stability condition (the Vakhitov-Kolokolov criterion), $dP/dK > 0$ [38,39], hence solitons cannot be stable in the model under the consideration. In the limit of $\alpha = 2$, Eq. (5) corresponds to the critical power for the self-focusing of the Gaussian beam in the framework of the usual two-dimensional nonlinear Schrödinger equation [39,40],

$$P_{\text{cr}} = \frac{3.77\lambda^2}{8\pi n_0 n_2}. \qquad (6)$$

We aim to solve Eq. (1) with the input in the form of the AGVBs, written in terms of the polar coordinates $\left(r, \phi\right)$:

$$u\left(r, \phi, z = 0\right) = A_0\text{Ai}\left(\frac{r_0 - r}{bw}\right)\exp\left(d\frac{r_0 - r}{bw}\right)\exp\left[-\frac{\left(r_0 - r\right)^2}{w^2}\right]\left(\frac{r}{w}\right)^l e^{il\phi}, \quad (7)$$

where $A_0$ is the overall amplitude, $A_i(\cdot)$ is the standard Airy function, $r_0$ is the radius of the primary Airy ring, $0 \le d < 1$ is the exponential truncation factor, $b$ is the distribution factor which determines the Airy-Gaussian width ratio, and integer $l$ is the topological charge of the optical vortex. The input beam is characterized by its total power (5), which takes the form of

$$P_{\text{in}} = 2\pi \int_0^\infty \left| u\left(r, z = 0\right) \right|^2 r \, dr \; . \tag{8}$$

It is relevant to mention that, similar to the usual nonlinear Schrödinger equation with $\alpha = 2$, the stationary version of FNS equation (1) admits solutions carrying integer vorticity, $l$. In this case, expression (4) is replaced by

$$u\left(x, y, z\right) = e^{iKz + il\phi} U_{K,l}\left(r\right), \tag{9}$$

with real function $U_{K,l}\left(r\right)$ satisfying equation

$$K U_{K,l} = -\frac{1}{2kw^{2-\alpha}} \left(-\nabla_r^2\right)^{\alpha/2} U_{K,l} + \frac{n_2 k}{n_0} U_{K,l}^3. \tag{10}$$

Here the fractional-diffraction operator, acting upon the function of radial coordinate $r$, is defined by an expression which follows from the substitution of ansatz (9) in Eqs. (2) and (3). Explicitly performing the inner integration with respect to the coordinate which is the angle between vectors $\left(k_x, k_y\right)$ and $\left(x - x', y - y'\right)$, one thus obtains

$$\left(-\nabla_r^2\right)^{\alpha/2} U_{K,l} = \frac{1}{2\pi} \int_0^\infty q^{\alpha+1} dq \int_0^\infty r' dr' \int_0^{2\pi} \cos(l\chi) d\chi$$
$$\times J_0\left(q\sqrt{r^2 + r'^2 - 2rr'\cos\chi}\right) U_{K,l}(r'), \tag{11}$$

where $J_0$ is the Bessel function. The possibility to construct the eigenstates in the form of Eqs. (9)-(11) implies that the angular momentum commutes with the fractal-diffraction operator in Eq. (1), similar to the fact that the total power (5) and the system's Hamiltonian commute with it.

Simulations of Eq. (1) were performed by means of the well-known split-step Fourier-transform method. In the simulations, we use constants which correspond, in physical units, to typical values (except for some specific cases) $n_0 = 1.45$, $\lambda = 532 \times 10^{-6}$ mm, $d = 0.2$, $r_0 = 1$ mm, $w = 1$ mm, and $n_2 = 2.6 \times 10^{-16}$ cm$^2$W$^{-1}$, while the distribution factor $b$ and the amplitude $A_0$ are varied as control parameters.

## 3. The propagation dynamics

Figure 1 displays results produced by the simulations for AGVBs initiated by input (4) with power $P_{\text{in}} = 6P_{\text{cr}}$, cf. Eqs. (5) and (6). As seen in Figs. 1(a) and 1(b1)-1(b4), under the action of the intrinsic focusing drive, these nonlinear beams autofocus after having passed distance measured in units of the Rayleigh length as $z =$ $0.5Z_R$, with the intensity (power density) contrast $I/I_0$, where $I$ is the beams' intensity along the propagation direction and $I_0$ is the initial intensity, attaining the maximum value of 21. Past the focal position, the center of the beams' transverse profile remains hollow, due to the usual effect of the intrinsic vorticity.

Figure 2 presents the three-dimensional plot of the intensity distribution of the same AGVBs in the course of the propagation, and the respective evolution of its peak intensity. Note that the transverse resolution of the numerical mesh, which is $3.4 \times 10^{-5}$ mm$^2$, is sufficient to capture the structure of the focused beam, whose minimum area is $9 \times 10^{-5}$ mm$^2$.

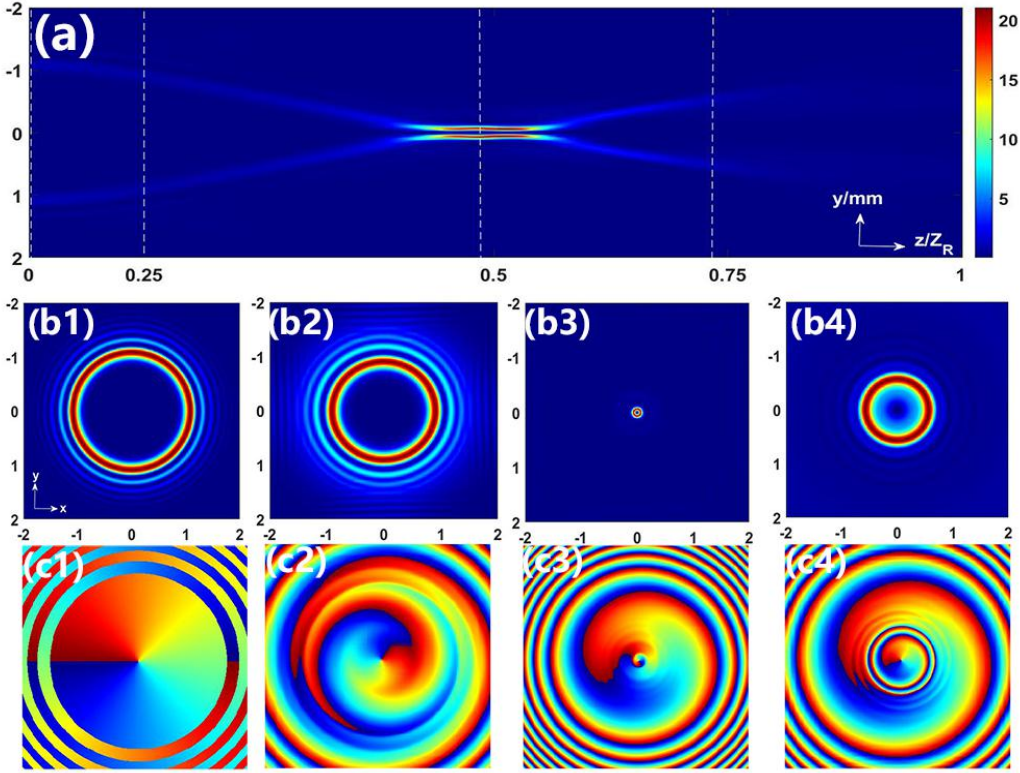

Fig. 1. (color online) Results of numerical simulations of Eq. (1) for input (7) in the form of the AGVBs with $b = 0.1, l = 1, \alpha = 1.5$, and $P_{\text{in}} = 6P_{\text{cr}}$. (a) The side view of the propagation of the beams. Panels (b1)-(b4) and (c1)-(c4) show intensities and phase distributions of the evolving AGVBs at positions denoted by vertical dashed white lines in panel (a).

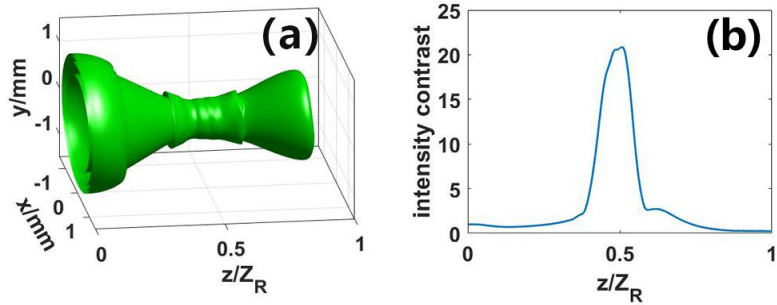

Fig. 2. (color online) The intensity isosurface (a) and intensity contrast (b) corresponding to the propagation of the AGVBs are shown in Fig. 1(a).

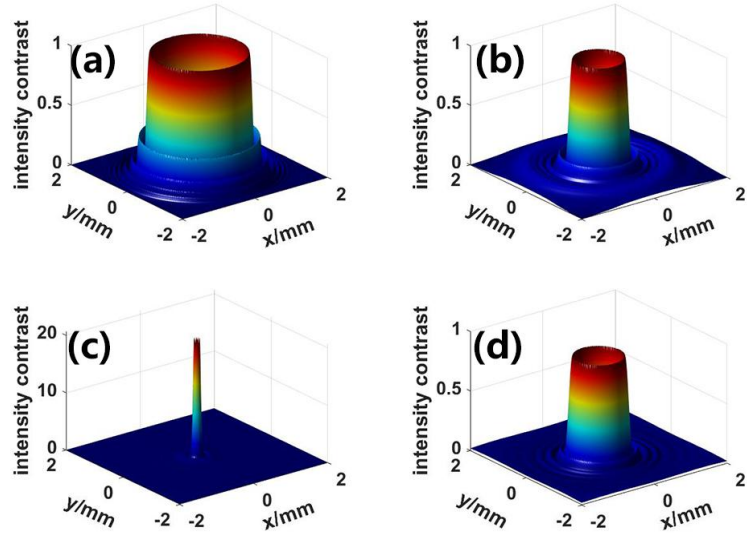

Fig. 3. (color online) The intensity contrasts of the AGVBs from Fig. 1 are shown in panels (a)-(d) at propagation distances $0, 0.25Z_R, 0.5Z_R$, and $0.75Z_R$, respectively.

Further, power profiles of the AGVBs introduced in Figs. 1 and 2 are shown at different propagation distances in Fig. 3. These plots illustrate in detail the autofocusing followed by expansion. In the present moderately nonlinear case, the beams self-focus only once, while the regime of stronger nonlinearity features multiple focusing events, as shown in Fig. 4 below.

The propagation dynamics of the AGVBs are strongly affected by the initial peak power $P_{in}$ defined in Eq. (8) and the distribution factor $b$ defined in Eq. (7). These effects are illustrated in Fig. 4 by means of dependencies of the beams' intensity contrast on the propagation distance, shown for different values of $P_{in}$ and $b$. Naturally, stronger nonlinearity, corresponding to larger $P_{in}$, leads to a shorter focusing distance. On the contrary, larger values of $b$, which imply narrower Airy rings in the input given by Eq. (7), produce a larger focusing distance, and a smaller peak intensity at the focal point.

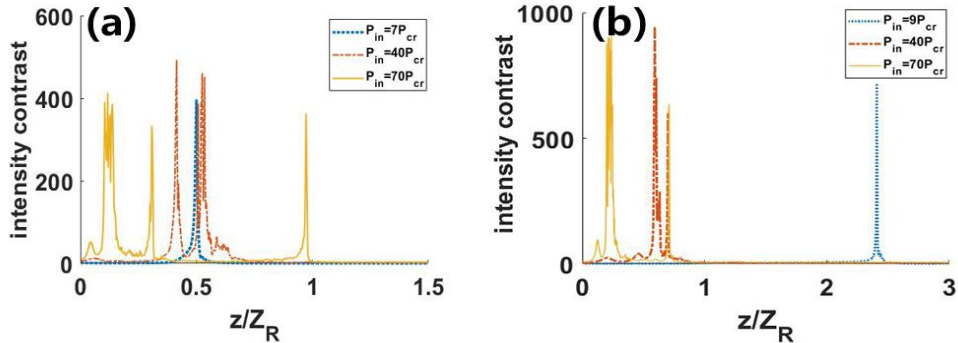

Fig. 4. (color online) The intensity contrast of the AGVBs vs. Propagation distance z, for three different values of input power (8), and two values of the distribution factor: b = 0.1 (a) and b =

0.3 (b). The other parameters as those in Fig. 1. In addition to the sharp self-compression, the strongly nonlinear axisymmetric vortex beams are subject to the azimuthal instability, which transforms them into necklace-shaped structures, see Fig. 9 below.

A specifically nonlinear effect, which takes place at larger values of $P_{in}$, is the second self-focusing event, which is followed, with the further increase of $P_{in}$, by a sequence of several focusing-defocusing cycles, see Figs. 4(a) and (b). In fact, the strong self-compression of such beams is a manifestation of the *quasi-collapse* (see Refs. [39] and [40]). Data produced by the numerical simulations, which were used to produce Fig. 4, make it possible to identify threshold values of $P_{in}$ at which the second autofocusing happens for the first time: these are $\left(P_{in}\right)^{(thr)}_{\text{second-focusing}} = 7P_{cr}$ for $b=0.1$ and $9P_{cr}$ for $b=0.3$, respectively.

At values of LI close to $\alpha = 1$, corresponding to strong fractality, another nonlinear effect is observed: the first focusing event is followed not by expansion, but by self-trapping of the beams into a *quasi-soliton*, that keeps a nearly constant transverse size with small-amplitude intrinsic vibrations, as shown by Fig. 5 for $\alpha = 1.1$ (a) and $\alpha = 1.2$ (b) and $P_{in} = 10P_{cr}$, other input parameters being the same as in Fig. 1. The quasi-soliton also keeps the initial vorticity, in the form of its internal angular momentum. However, for LI closer to the ordinary value of $\alpha = 2$, such as $\alpha = 1.5$, the beams with the same power, $P_{in} = 10P_{cr}$, expand after the self-focusing stage, without forming a quasi-soliton (not shown here in detail). Thus, the creation of the quasi-soliton is a specific manifestation of the interplay of the relatively strong fractality and nonlinearity.

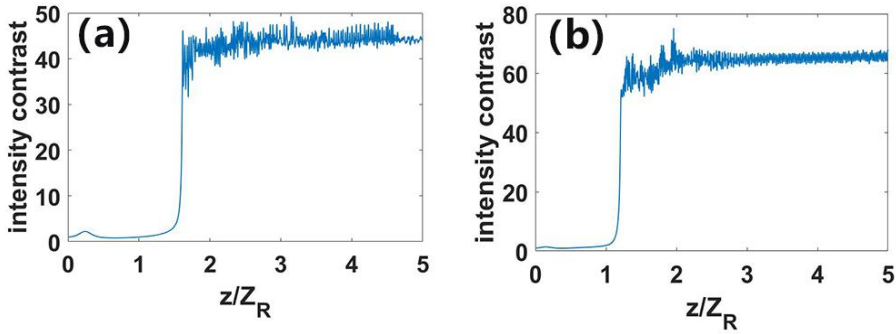

Fig. 5. (color online) The intensity contrast of the vortex beam vs. propagation distance $z$, with the same parameters as those in Fig. 1, except for $P_{in} = 10P_{cr}$, at relatively small values of LI: $\alpha = 1.1$ (a) and $\alpha = 1.2$ (b). In this case, the input forms a quasi-soliton with small intrinsic oscillations.

Next, in Fig. 6 we aim to address effects of the change of LI. It is observed that the increase of LI makes the onset of the autofocusing much faster, the focusing distance being much shorter in Fig. 6(b) than that in Fig. 6(a). On the other hand, the intensity contrast in the focal plane is essentially higher at smaller LI [in Fig. 6(a1), it is larger than that in Fig. 6(b1) by a factor $\approx 4$]. In line with this observation, smaller LI makes the focal profile narrower. The difference is explained by the fact that the longer autofocusing distance in panel (a) makes it possible for the AGVBs to achieve stronger self-compression.

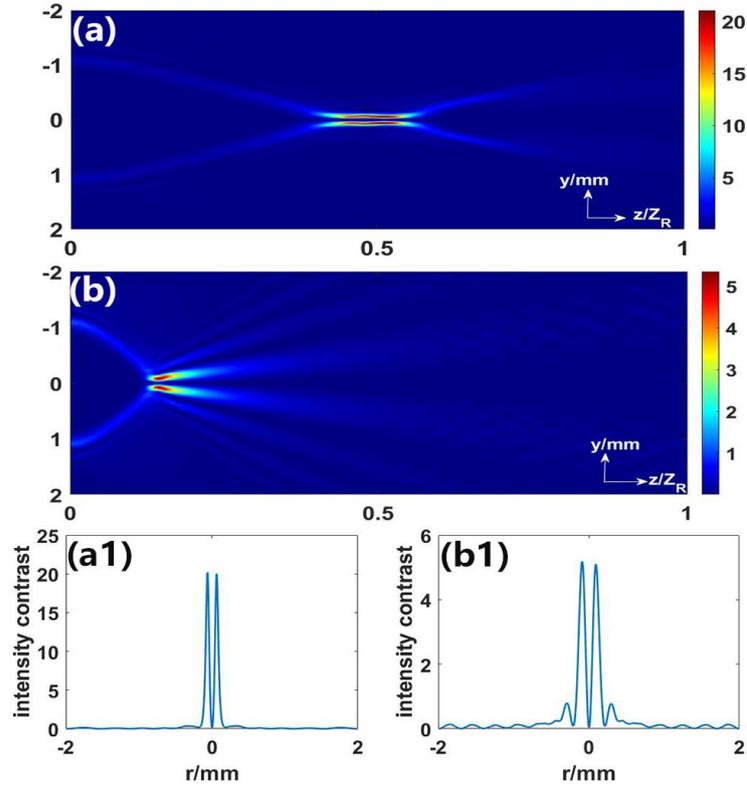

Fig. 6. (color online) Side views of autofocusing nonlinear AGVBs at two different values of LI: $\alpha = 1.5$ (a) and $\alpha = 1.8$ (b). Panels (a1) and (b1) show intensity contrasts in the focal planes for the same cases. Other parameters are $l = 1$, $b = 0.1$, and $P_{in} = 6P_{cr}$.

The dependence of the focal intensity and the autofocusing distance on LI are presented, in a systematic form, in Fig. 7. In agreement with Fig. 6, it shows a monotonous dependence of the autofocusing distance on $\alpha$, but non-monotonous variation of the intensity contrast following the increase of $\alpha$. Accordingly, the intensity attains a maximum in Fig. 7(a) at $\alpha \approx 1.4$ (a similar result was very recently reported in Ref. [35]). Then, it rapidly decreases with the increase of $\alpha$, dropping to a minimum value in the limit of $\alpha = 2$, which corresponds to the ordinary diffraction in the two-dimensional space. Thus, the focal intensity, attained by the AGVBs under the action of the fractional diffraction, is *always higher* than that provided by the ordinary diffraction.

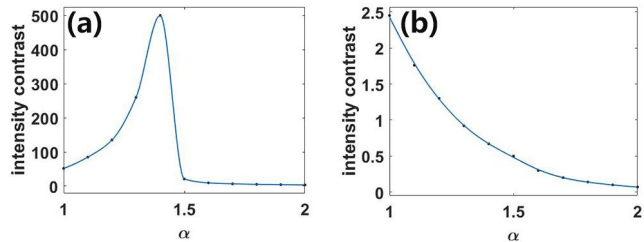

Fig. 7. (color online) The intensity contrast (a) and the autofocusing distance (b) of the nonlinear AGVBs vs. LI $\alpha$. Other parameters are the same as those in Fig. 6.

The next objective is to explore the autofocusing dynamics of the AGVBs with different vorticities $l$. As shown by the numerical results collected in Fig. 8, larger values of $l$ lead to the increase of the number of rings in the original intensity profile, a wider hollow channel in the focused state, and the decrease of the peak intensity of the beams. On the other hand, the focusing distance is almost independent of $l$.

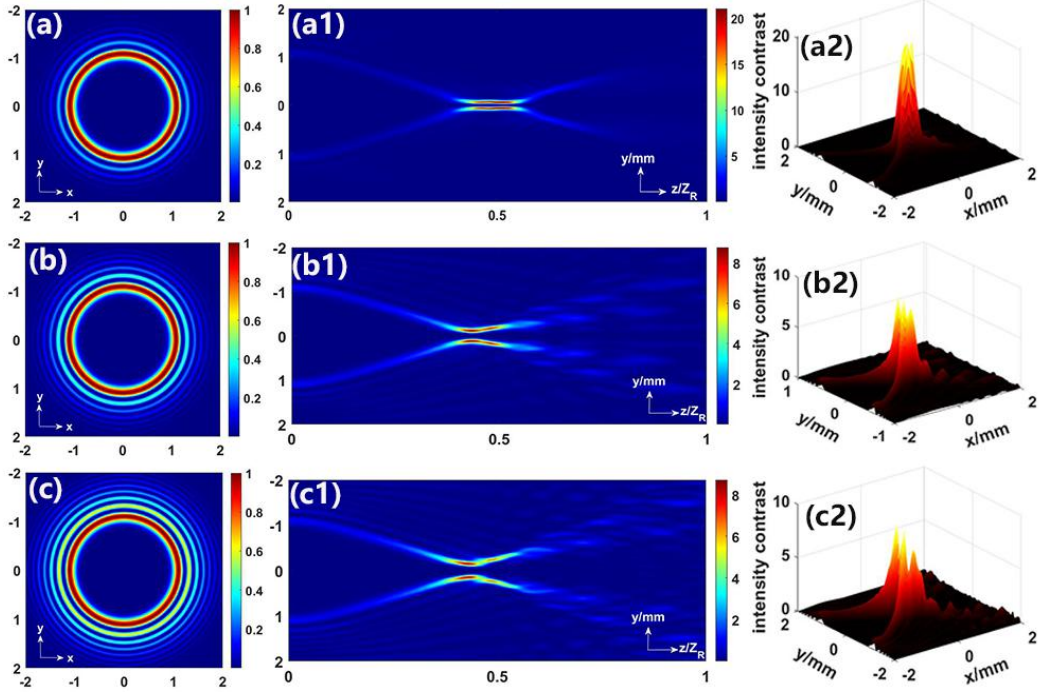

Fig. 8. (color online) The evolution of AGVBs with the distribution factor $b = 0.1$ and different vorticities: $l = 1$ (a)-(a2), $l = 2$ (b)-(b2), and $l = 3$ (c)-(c2). Panels (a)-(c) present the intensity contrasts of the respective inputs. (a1)-(c1): Side views of the simulated evolution. (a2)-(c2): The intensity contrasts of the AGVBs vs. the propagation distance. The other parameters are the same as those in Fig. 1.

Another conspicuous dynamical effect revealed by the simulations is a possibility of splitting of the vortex into a necklace-shaped rotating cluster. As an example, we consider the propagation of the AGVBs with $P_{in} = 30P_{cr}$. The respective intensity distributions are plotted in Figs. 9(a1,a2) and 9(b1,b2) at different propagation distances, for $P_{in} = 20P_{cr}$ and $P_{in} = 30P_{cr}$, respectively. The figure clearly demonstrates that the azimuthal modulational instability splits the initially axisymmetric pattern into a set of four bright spots, which separate in the course of the propagation. As shown in Figs. 9(b1) and 9(b2), we can see that transverse structure of the beams maintains a set of four bright spots at different propagation distances, $z \approx 0.42Z_R$ and $0.67Z_R$ with $P_{in} = 30P_{cr}$, respectively. It is also observed that the transverse structure rotates in the course of the propagation.

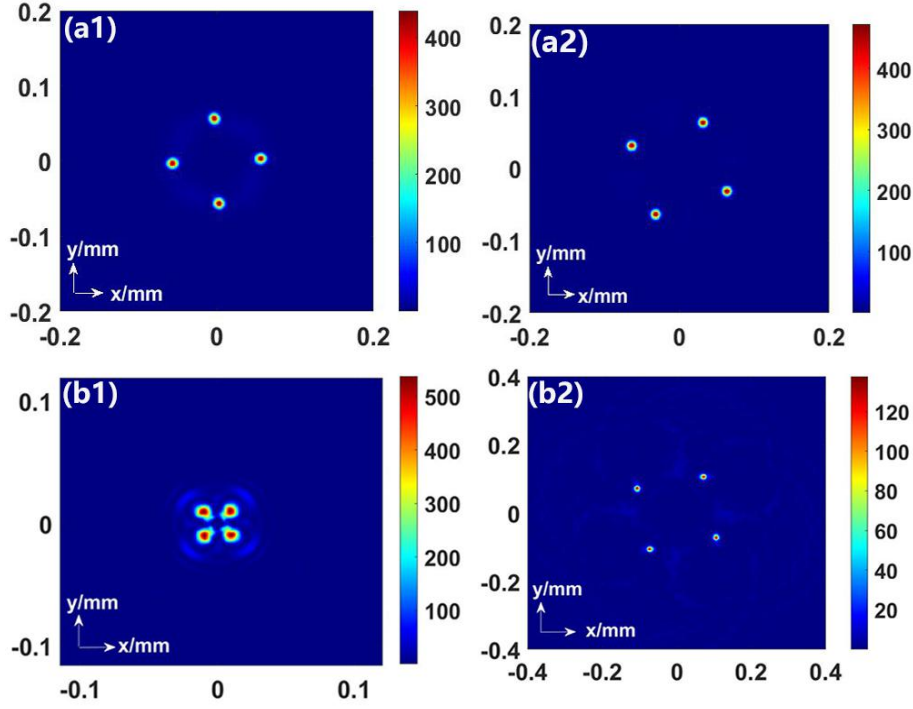

Splitting of the vortex beams with a thin radial structure [such as one observed in Figs. 1(b4), 2(a), and 3(d)] into a radially confined cluster of fragments, under the action of the azimuthal modulational instability, may be approximated by an effectively one-dimensional, although nonlocal, equation written in the azimuthal direction, while the radial structure is eliminated. To this end, the two-dimensional transverse structure of the beams is adopted in the form of

$$u\left(r, \phi, z\right) \approx \frac{1}{\pi^{1/4}\sqrt{\sigma}}\exp\left(-\frac{\left(r - r_0\right)^2}{2\sigma^2}\right)v\left(\phi, z\right), \qquad (12)$$

where $r_0$ is the radius of the thin structure, and $\sigma \ll r_0$ is its small thickness. Then, substitution of approximation (12) in Eq. (1) and straightforward averaging in the radial direction leads to the equation for $v(\phi, z)$:

$$i\frac{\partial v}{\partial z} - \frac{r_0}{4\pi k w^{2-\alpha}} \int_0^\infty q^{\alpha+1} dq \int_0^{2\pi} d\phi' J_0\left(2r_1 q \sin\left(\frac{\phi-\phi'}{2}\right)\right) v(\phi')$$

$$+ \frac{n_2 k}{n_0 \sqrt{2\pi\sigma}} |v|^2\, v = 0. \tag{13}$$

Further analysis of the azimuthal equation (13) will be presented elsewhere.

To further investigate the propagation characteristics of the AGVBs under the action of the fractional diffraction, we introduce the Poynting vector and the angular momentum of the underlying electromagnetic field. To this end, the vector potential of the field is introduced as $\vec{A} = \vec{e}_x u(x, y, z) \exp(ikz)$, being directed along the $x$ direction (as denoted by unit vector $\vec{e}_x$). Then, the time-averaged Poynting vector of the field can be expressed as (see, e.g., Ref. [41])

$$\left\langle \vec{S} \right\rangle = \frac{c}{4\pi} \left\langle \vec{E} \times \vec{B} \right\rangle = \frac{c\omega}{8\pi} \left[ i\left(u\nabla_\perp u^* - u^*\nabla_\perp u\right) + 2k|u|^2\, \vec{e}_z \right], \tag{14}$$

where $c$ is the speed of light in vacuum, $\vec{E}$ and $\vec{B}$ are the electric and the magnetic components of the field, respectively, and $\nabla_\perp = \frac{\partial}{\partial x}\vec{e}_x + \frac{\partial}{\partial y}\vec{e}_y$, $\vec{e}_y$ and $\vec{e}_z$ being the unit vector along the $y$ direction and the $z$ direction, and * stands for the complex conjugate.

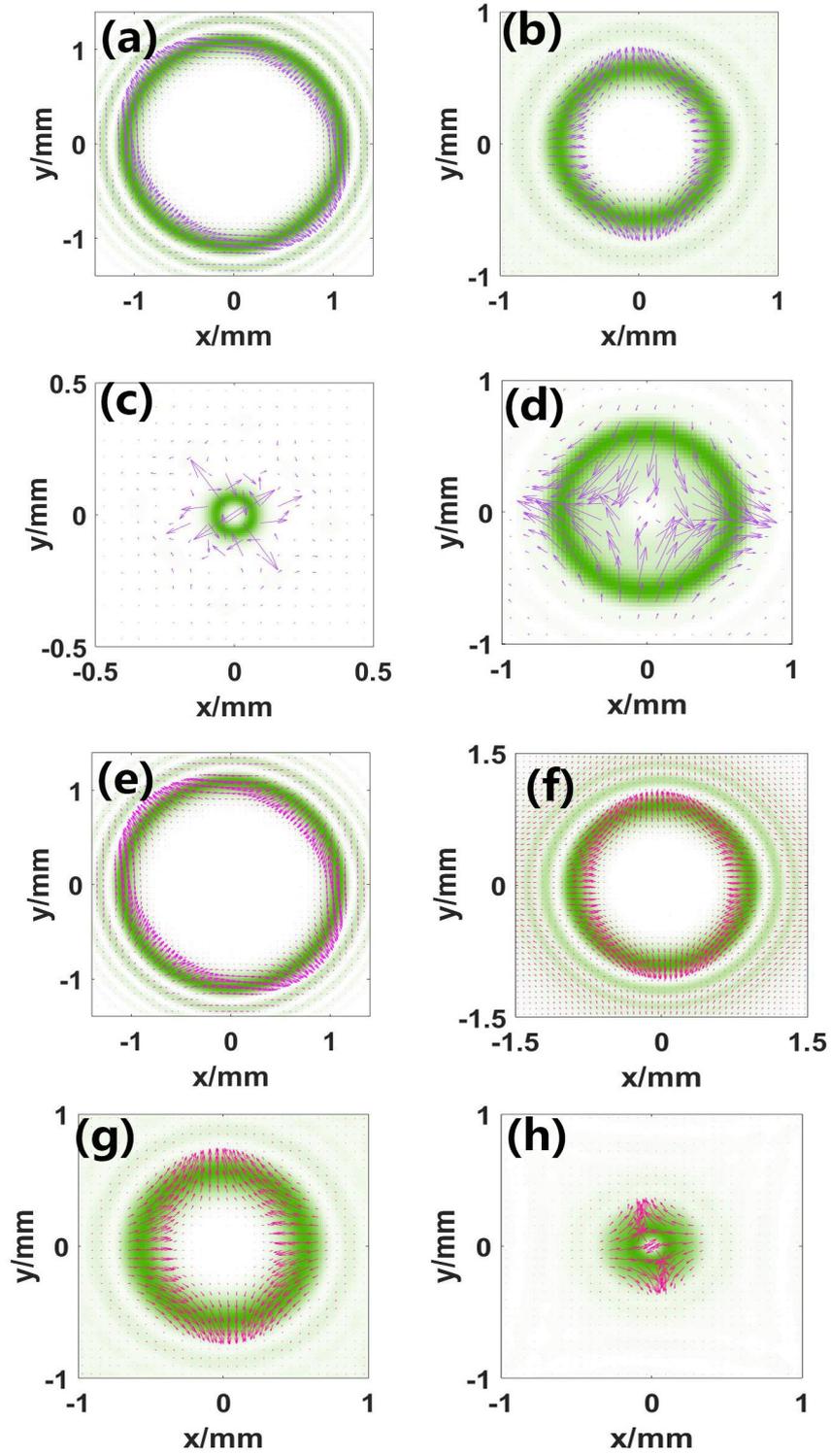

Fig. 10. (a)-(d): The transverse energy flow, determined by the longitudinal component of the Poynting vector (14) of the AGVBs in the NLS medium. (e)-(h): The same in the linear medium (free space). The flow is presented by fields of arrows at different propagation distances corresponding to Figs. 1(b1)-1(b4).

The direction and magnitude of the local power flow in the transverse plane ($x$,$y$)

are determined by the (small) longitudinal component of the Poynting vector, the distributions of which, corresponding to the AGVBs from Fig. 1, are represented in Fig. 10 at different values of the propagation distance. Naturally, the power flow originally converges towards the focal ring, and then reverses. Note also that, in the initial main ring, the power flow tends to go from the x direction to the y direction, and turns around outside of the ring. During the evolution, the direction of the energy flow in the AGVBs changes. As a result, the beams' energy flows chiefly from the center to periphery in Fig. 10(c). Past the focal plane, as shown in Fig. 10(d), the energy flows from the upper and lower sides into the center, while the flows from the left and the right sides are directed outward.

The initial flow distributions of the AGVBs are identical in the free space and NLS medium. As seen in Figs. 10(a) and 10(e), the respective power flow tends to rotate from the x to y direction, and turns around outside of the ring. In the course of the evolution in the free space, the energy flow directions in Fig. 10(f) remain the same as in Fig. 10(e). Note that, in the absence of the nonlinearity, the energy flows to the center slower than in NLS medium. The same trend is seen in Figs. 10(g) and 10(h). As a result, no autofocusing event takes place in the linear medium before $z = 0.75Z_R$. The fact that the energy flow in the linear medium still converges to the center is completely different from what is observed in Figs. 10(c) and 10(d) in the nonlinear medium.

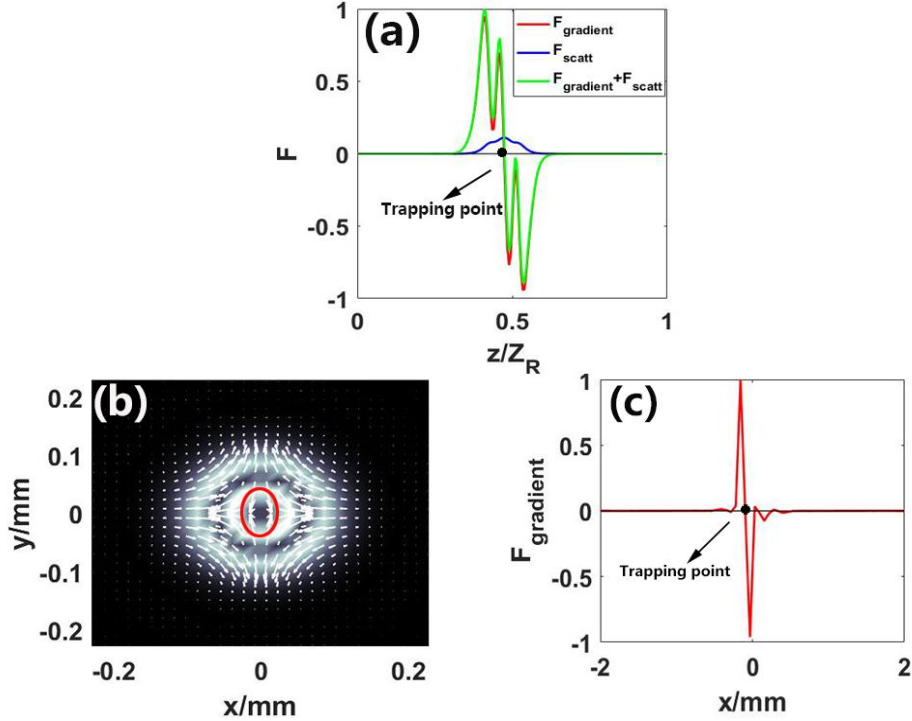

Fig. 11. (color online) (a) The distribution of the z-component of the maximum radiation force in the AGVBs along the z- axis, in arbitrary units. (b) The distribution of the z-component of the maximum gradient force as functions of x and y in the focal plane. (c) The distribution of the z-component of the maximum gradient force in the focal plane. The forces are calculated as per Eqs. (15) and (16). All the parameters are the same as those in Fig. 1, while the size of the probe

particle, onto which the forces are acting, is taken as $\mu_0$=0.5 nm, in physical units.

The radiation force which determines the capturing capability of the electromagnetic field, includes two components, *viz.*, the gradient and the scattering ones [42]:

$$\vec{F}_{grad}\left(x,y,z\right) = \frac{2\pi\mu_0^3 n_3}{c}\left(\frac{m^2-1}{m^2+2}\right)\nabla I\left(x,y,z\right), \qquad (15)$$

$$\vec{F}_{scatt}\left(x,y,z\right) = \frac{8\pi k^4 \mu_0^6 n_3}{3c}\left(\frac{m^2-1}{m^2+2}\right)I\left(x,y,z\right)\vec{e}_z, \qquad (16)$$

where $m = n_1/n_3$ is the relative refractive index of the probe nano-particle (here, we assume $n_1 = 1.5$ and $n_3 = 1$ for the medium), $\mu_0$ is its radius, the intensity is $I\left(x,y,z\right) = \frac{1}{2}cn_3\varepsilon_0\left|\vec{u}\left(x,y,z\right)\right|^2$, and $\nabla$ is the full three-dimensional gradient.

The calculation, based on Eqs. (15) and (16), makes it possible to predict an equilibrium position at which a probe nanoparticle may be captured by the AGVBs in the autofocusing plane, due to cancellation of the gradient and the scattering forces. The equilibrium position is shown by the black dot in Figs. 11(a,c), and by a small red circle in Fig. 11(b). Thus, the AGVBs may be used for the design of effective optical tweezers.

## 4. Conclusion

We have reported basic results for autofocusing of AGVBs (Airy-Gauss vortex beams) propagating under the actions of the fractional diffraction and cubic self-attraction, which helps to initiate the autofocusing of the beams. While the focusing length monotonously decreases as the fractional LI (Lévy index) approaches $\alpha = 2$, i.e., the normal 2D diffraction, the peak intensity at the focal point attains a strong maximum at $\alpha \approx 1.4$. The propagation regimes feature the single autofocusing event. In contrast with that, the nonlinearity, if it is strong enough, gives rise to multiple autofocusing events, as well as to splitting of axisymmetric beams into a set of bright spots. In the case of strong fractality (corresponding to $\alpha$ close to 1), after the first instance of autofocusing the nonlinear beam may self-trap into a vortical quasi-soliton with intrinsic oscillations. Finally, the ability of the electromagnetic field carrying the AGVBs to capture a nano-particle is explored. An equilibrium position, provided by the cancellation of radiation forces, is identified.

The predicted propagation regimes may be used in optical data-processing systems, as well as for the design of optical trapping and transfer of nanoparticles.

## FUNDING INFORMATION

This work was supported by the National Natural Science Foundation of China (11775083, 11374108, 61675001, 11947103, 12004081), Science and Technology Program of Guangzhou (No. 2019050001), and the Israel Science Foundation (grant

No. 1286/17).

**CONFLICT OF INTEREST**

We declare that we have no conflict of interest.